\documentclass[prl,twocolumn,superscriptaddress,calc,epsfig]{revtex4-1}
\usepackage{graphics}
\usepackage{amsfonts}
\usepackage{amsmath}
\usepackage{epsfig}

\newif\ifdraft \drafttrue

\begin{document}

\title{Universal Spin Transport in a Strongly Interacting Fermi Gas}
\catcode`\ä = \active \catcode`\ö = \active \catcode`\ü = \active
\catcode`\Ä = \active \catcode`\Ö = \active \catcode`\Ü = \active
\catcode`\ß = \active \catcode`\é = \active \catcode`\è = \active
\catcode`\ë = \active \catcode`\ô = \active \catcode`\ê = \active
\catcode`\ø = \active \catcode`\ò = \active \catcode`\í = \active
\catcode`\Ó = \active \catcode`\ú = \active \catcode`\á = \active
\catcode`\ã = \active \catcode`\à = \active
\defä{\"a} \defö{\"o} \defü{\"u} \defÄ{\"A} \defÖ{\"O} \defÜ{\"U} \defß{\ss} \defé{\'{e}}
\defè{\`{e}} \defë{\"{e}} \defô{\^{o}} \defê{\^{e}} \defø{\o} \defò{\`{o}} \defí{\'{i}}
\defÓ{\'{O}} \defú{\'{u}} \defá{\'{a}} \defã{\~{a}}\defà{\`{a}}

\newcommand{\mitcua}{Department of Physics, MIT-Harvard Center for Ultracold Atoms, and Research Laboratory of Electronics,
        MIT, Cambridge, Massachusetts 02139, USA}
\newcommand{\lens}{LENS and Dipartimento di Fisica, Università di Firenze, and INO-CNR, 50019 Sesto Fiorentino, Italy}

\author{Ariel Sommer}
\author{Mark Ku}
\affiliation{\mitcua}
\author{Giacomo Roati}
\affiliation{\lens}
\author{Martin W. Zwierlein}
\affiliation{\mitcua}

\newcommand{\li}{$^6$Li}
\newcommand{\na}{$^{23}$Na}
\newcommand{\cs}{$^{133}$Cs}
\newcommand{\kk}{$^{40}$K}
\newcommand{\rb}{$^{87}$Rb}
\newcommand{\vect}[1]{\mathbf #1}
\newcommand{\g}{g^{(2)}}
\newcommand{\one}{$\left|\uparrow\right>$}
\newcommand{\two}{$\left|\downarrow\right>$}
\newcommand{\V}{V_{12}}
\newcommand{\GammaSD}{\ensuremath{\Gamma_{\mathrm{sd}}}}
\newcommand{\GammaColl}{\ensuremath{\Gamma_{\mathrm{coll}}}}
\newcommand{\normSD}{\ensuremath{\tilde{\Gamma}_{\mathrm{sd}}}}
\newcommand{\normDs}{\ensuremath{\tilde{D}_{\mathrm{s}}}}
\newcommand{\Ds}{\ensuremath{D_{\mathrm{s}}}}
\newcommand{\chis}{\ensuremath{\chi_{\mathrm{s}}}}
\newcommand{\normCs}{\ensuremath{\tilde{\chi}_{\mathrm{s}}}}
\newcommand{\ag}{\ensuremath{\alpha_{\mathrm{\Gamma}}}}
\newcommand{\ad}{\ensuremath{\alpha_{\mathrm{D}}}}
\newcommand{\kfa}{\frac{1}{k_F a}}
\newcommand{\resField}{\ensuremath{834 \,\rm G}}
\newcommand{\kickField}{\ensuremath{50 \,\rm G}}
\newcommand{\prepField}{\ensuremath{300 \,\rm G}}
\newcommand{\zup}{\ensuremath{z_\uparrow}}
\newcommand{\zdown}{\ensuremath{z_\downarrow}}
\newcommand{\vzup}{\ensuremath{v_{z\uparrow}}}
\newcommand{\vzdown}{\ensuremath{v_{z\downarrow}}}

\begin{abstract}
Transport of fermions is central in many fields of physics. Electron transport runs modern technology, defining states of matter such as superconductors and insulators, and electron spin, rather than charge, is being explored as a new carrier of information~\cite{wolf01spin}. Neutrino transport energizes supernova explosions following the collapse of a dying star~\cite{burr90neut}, and hydrodynamic transport of the quark-gluon plasma governed the expansion of the early Universe~\cite{scha09perfect}. However, our understanding of non-equilibrium dynamics in such strongly interacting fermionic matter is still limited. Ultracold gases of fermionic atoms realize a pristine model for such systems and can be studied in real time with the precision of atomic physics~\cite{zwie05vort,vare06fermi}. It has been established that even above the superfluid transition such gases flow as an almost perfect fluid with very low viscosity~\cite{scha09perfect,cao10univ}
when interactions are tuned to a scattering resonance. However, here we show that spin currents, as opposed to mass currents, are maximally damped, and that interactions can be strong enough to reverse spin currents, with opposite spin components reflecting off each other. We determine the spin drag coefficient, the spin diffusivity, and the spin susceptibility, as a function of temperature on resonance and show that they obey universal laws at high temperatures. At low temperatures, the spin diffusivity approaches a minimum value set by $\hbar/m$, the quantum limit of diffusion, where $\hbar$ is the reduced Planck's constant and $m$ the atomic mass. For repulsive interactions, our measurements appear to exclude a metastable ferromagnetic state~\cite{jo09ferro, stri09dens, duin10spin}.

\end{abstract}
\maketitle

Understanding the transport of spin, as opposed to the transport of charge, is of high interest for the novel field of spintronics~\cite{wolf01spin}. While charge currents are unaffected by electron-electron scattering due to momentum conservation, spin currents will intrinsically damp due to collisions between opposite spin electrons, as their relative momentum is not conserved. This phenomenon is known as spin drag~\cite{dami00theo, webe05obse}. It is expected to contribute significantly to the damping of spin currents in doped semiconductors~\cite{dami02coul}. The random collision events also lead to spin diffusion, the tendency for spin currents to flow such as to even out spatial gradients in the spin density, which has been studied in high-temperature superconductors~\cite{gedi03diff} and in liquid $^3$He-$^4$He solutions~\cite{garw59self, ande66ther}.

Creating spin currents poses a major challenge in electronic systems where mobile spins are scattered by their environment and by each other. However, in ultracold atoms we have the freedom to first prepare an essentially non-interacting spin mixture, separate atoms spatially via magnetic field gradients, and only then induce strong interactions. Past observations of spin currents in ultracold Fermi gases~\cite{dema02spin, du08obse} were made in the weakly-interacting regime. Here we access the regime near a Feshbach resonance~\cite{vare06fermi}, where interactions are as strong as allowed by quantum mechanics (the unitarity limit). We measure spin transport properties, the spin drag coefficient \GammaSD{} and the spin diffusivity \Ds{}, of a strongly interacting Fermi gas composed of an equal number of atoms in two different spin states. In the strongly-interacting regime, spin drag is expected to reach a universal maximum value, and spin diffusion is expected to reach a universal minimum.

The universal behaviour of the spin transport coefficients of a Fermi gas can be estimated on general grounds.
At the Feshbach resonance, the scattering cross-section $\sigma$ between atoms of opposite spin is given by the square of the matter wavelength, in the degenerate regime $\sigma \sim 1/k_F^2$, where $k_F = (6\pi^2 n)^{1/3}$ is the Fermi wavevector and $n$ is the density of atoms in each spin state. The mean free path between collisions is thus $l = 1/n\sigma \sim 1/k_F$ or about one interparticle spacing, the smallest possible mean free path in a gas. The average speed $v$ of atoms is on the order of the Fermi velocity $\hbar k_F/m$. In estimating the spin diffusivity $\Ds \approx v l$ the density-dependent factors cancel, giving $\Ds \approx \hbar / m$. This value for $\Ds$ represents a universal quantum limit to spin diffusivity in Fermi gases. Away from resonance, the scattering cross section decreases, increasing \Ds. For temperatures $T$ much greater than the Fermi temperature $T_F = \hbar^2 k_F^2/2 m k_B$, the scattering cross section will be given by the square of the thermal de Broglie wavelength and thus decrease as $\sigma \propto 1/T$, while the velocity will increase as $v\propto \sqrt{T}$, causing \Ds{} to increase as $\Ds \propto T^{3/2}$. Finally, in a degenerate Fermi gas the average velocity will remain on the order of the Fermi velocity, but the effective scattering cross section will scale as $\sigma \propto T^{2}$ due to Pauli blocking, causing \Ds{} to increase like $T^{-2}$ as the temperature is lowered. For a Fermi gas, we thus expect the minimum \Ds{} to occur at temperatures near the Fermi temperature. Correspondingly, the coefficient \GammaSD{} characterizing spin drag is expected to reach a universal
maximum value on resonance and for temperatures near the Fermi temperature, given by the Fermi rate $E_F/\hbar$, where $E_F = \hbar^2 k_F^2/2m$ is the Fermi energy.
\begin{figure}[ht]
    \begin{center}
    \includegraphics[width=89mm]{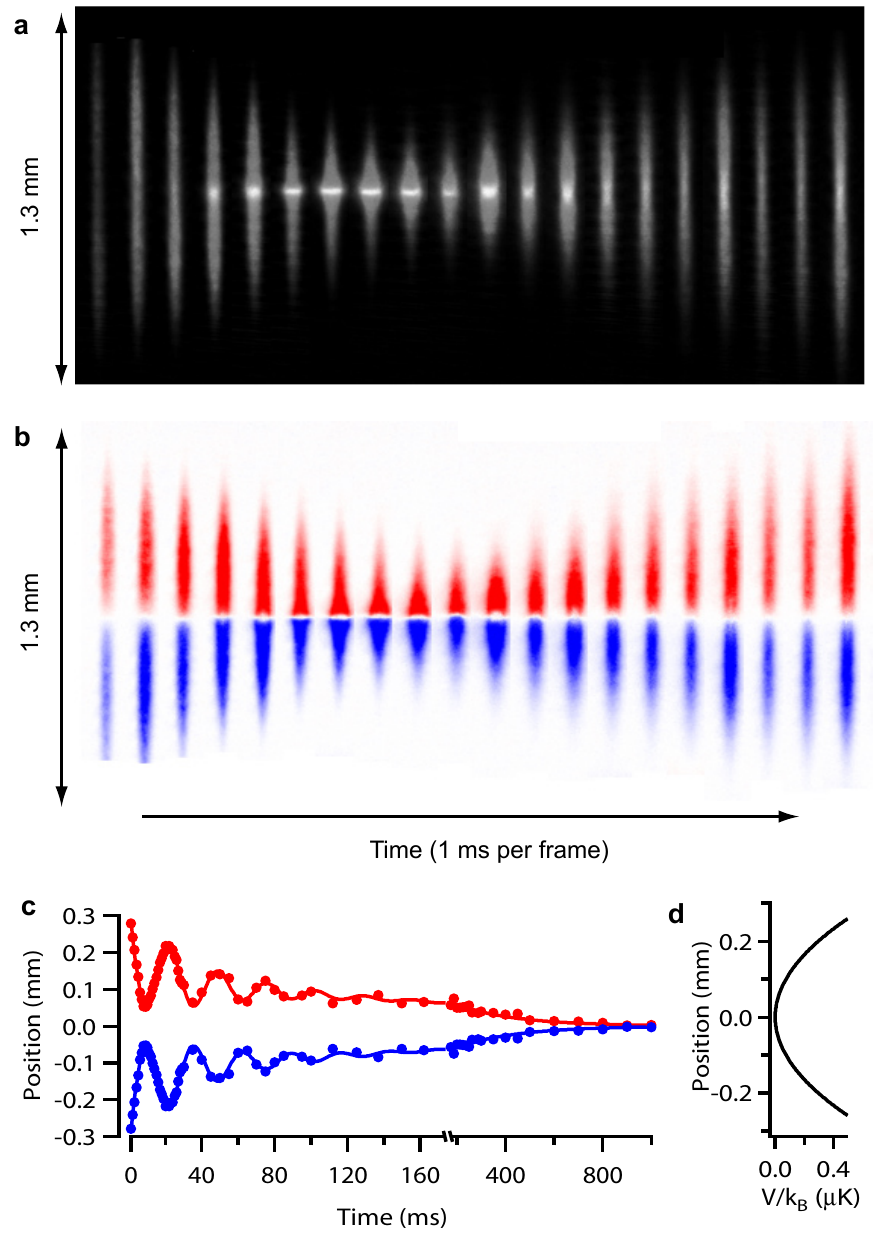}
    \caption[Title]{{\bf Observation spin current reversal in a resonant collision between two oppositely spin-polarized clouds of fermions.} ({\bf a}) shows the total column density and ({\bf b}) the difference in column densities of the two clouds
    (red: spin up, blue: spin down), in 1 ms intervals during the first 20 ms after the magnetic field is set to the Feshbach resonance at 834 G. The collision leads to the formation of a high-density interface between the two spin states. ({\bf c}) The separation between the centers of mass of the two spin states initially oscillates at a frequency of 1.63(2) $\omega_z$, where $\omega_z = 2\pi \times 22.8$ Hz is the trap frequency in the axial direction (see Sup. Info. for discussion). Even after half a second, there is still substantial spin separation. The diffusion time indicates a diffusivity on the order of $\hbar/m$. ({\bf d}) Shows the harmonic trapping potential along the axis of symmetry.}
    \label{fig:bounces}
    \end{center}
\end{figure}

In the experiment, we prepare an equal mixture of the two lowest hyperfine states (``spin up'' and ``spin down'') of fermionic $^6$Li in a cylindrically symmetric atom trap~\cite{zwie05vort,vare06fermi}. The confinement along the axis of symmetry is harmonic, with frequency $\omega_z$. We separate the two spin components along the axis of symmetry of the trap (see Methods Summary) and turn on strong interactions between unequal spins by quickly ramping the magnetic field to a Feshbach resonance located at 834 G. The confining potential of the trap forces the two clouds of opposite-spin atoms to propagate towards each other, establishing a spin current. Measurements are made by selectively imaging the two spin components.

Figure 1 shows the collision between the two spin domains on resonance. The clouds bounce off each other and essentially completely repel. Due to the axial trapping potential, the clouds return after the collision, and we observe several oscillations in the displacement $d = \langle z_{\uparrow}\rangle-\langle z_{\downarrow}\rangle$, where $\langle z_{\uparrow(\downarrow)}\rangle$ is the center of mass of the spin up (down) cloud. After the oscillations have decayed, the displacement decreases to zero monotonically, on a timescale on the order of one second, an extremely long time compared to the trapping period (44 ms).
The underlying explanation for spin current reversal and the slow relaxation can be found in the extremely short mean free path and the high collision rate between opposite-spin atoms at unitarity. According to the above estimate, the spin diffusivity is approximately $\hbar/m$, which for $^6$Li is (100 $\mu$m)$^2/$s. The atom clouds in the experiment have a length on the order of 100 $\mu$m, and it takes them on the order of a second to diffuse through each other. So we are indeed observing quantum-limited spin diffusion. The initial bounces will occur when the mean free path of a spin up atom in the spin down cloud is smaller than the spin down cloud size, i.e. when the mixture is hydrodynamic. Instead of quickly diffusing into the spin down region, it is then more likely that the spin up atom is scattered back into the spin up region where it can propagate ballistically. Indeed, we see bounces occur already for interaction strengths away from the Feshbach resonance where the mean free path is on the order of the axial cloud size (see Supplementary Information).

The relaxation dynamics close to equilibrium give direct access to the spin transport coefficients. The spin drag coefficient \GammaSD{} is defined as the rate of momentum transfer between opposite spin atoms~\cite{dami00theo}, and is therefore related to the collision rate. The relaxation of the displacement $d$ near equilibrium then follows the differential equation~\cite{vich99coll}
$$\GammaSD \dot{d} + \omega_z^2 d = 0.$$
Fitting an exponential with decay time $\tau$ to the displacement gives the spin drag coefficient as $\GammaSD = \omega_z^2 \tau$.

The spin drag coefficient is found to be greatest on resonance, and thus spin conduction is slowest (see Supplementary Information). On dimensional grounds, $\GammaSD$ must be given by a function of the reduced temperature $T/T_F$ times the Fermi rate $E_F/\hbar$. At high temperatures, we expect the spin drag coefficient to obey a universal scaling $\GammaSD \sim n \sigma v \sim T^{-1/2}$. In fig. 2 we show the spin drag coefficient as a function of $T/T_F$. We observe $T^{-1/2}$ scaling for $T/T_F > 2$, finding $\GammaSD = 0.16(1) \frac{E_F}{\hbar}(T/T_F)^{-1/2}$. At lower temperatures, we observe a crossover from classical to non-classical behavior: the spin drag coefficient reaches a maximum of approximately $0.1 E_F/\hbar$ near the Fermi temperature. We interpret this saturation of the spin drag coefficient as a consequence of Fermi statistics, as $\sigma$ and $v$ approach constant values~\cite{dema02spin} determined by the Fermi wavevector $k_F$. The maximum spin drag coefficient corresponds to a minimum spin conductivity $\sigma_s = \frac{n}{m\GammaSD}$ on the order of $k_F/\hbar$. This is the slowest spin conduction possible in three dimensions in the absence of localization.
\begin{figure}[ht]
    \begin{center}
    \includegraphics[width=89mm]{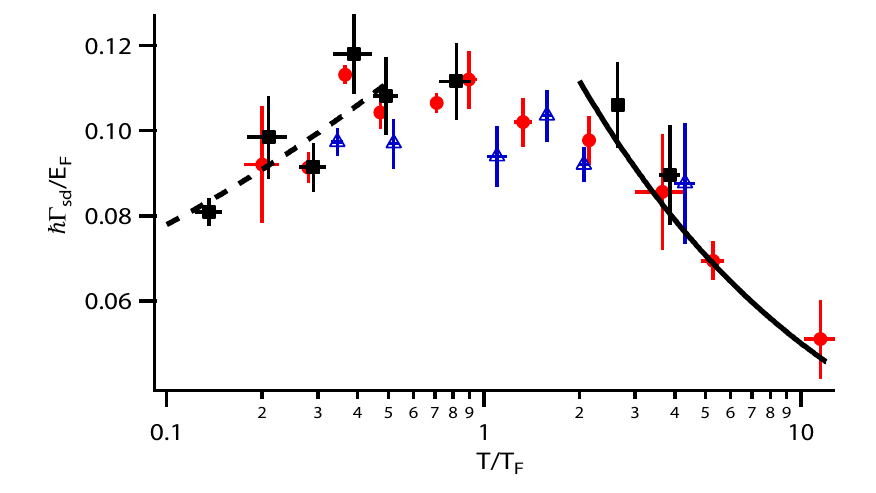}
    \caption[Title]{{\bf Spin drag coefficient of a trapped Fermi gas.} Spin drag coefficient for resonant interactions, as a function of the dimensionless temperature. The temperature is normalized by the maximum Fermi temperature, defined in terms of the maximum density of a single spin state in the system. We find agreement between measurements taken at three different values of the axial trapping frequency $\omega_z = 2\pi \times 22.8\,{\rm Hz}$ (red circles), $\omega_z = 2\pi\times 37.5\,{\rm Hz}$ (blue triangles), $\omega_z = 2\pi\times 11.2\, {\rm Hz}$ (black squares). The data for $T/T_F > 2$ fits to a $T^{-1/2}$ law (solid line). Dashed line: a power law fit with an unconstrained exponent for $T/T_F < 0.5$ to show the trend. The spin drag coefficient is obtained from the decay rate of the spin separation and normalized by the square of the trap frequency and by the maximum Fermi energy. The error bars are statistical.}
    \label{fig:spindrag}
    \end{center}
\end{figure}

At low temperatures, the spin drag coefficient decreases with decreasing temperature. No theoretical prediction for the spin drag coefficient exists in this regime. For highly-imbalanced spin populations, a decrease in the spin drag coefficient is expected due to Pauli blocking~\cite{bruu08coll}, which reduces the collision rate in the gas by a factor of $(T/T_F)^2$ at sufficiently low temperatures. However, for systems with equal spin populations, it was shown that pairing correlations enhance the effective collision rate for collective excitations, leading to a collision rate that increases dramatically as the temperature is lowered~\cite{ried08coll}. The observed reduction of \GammaSD{} at low temperatures contrasts with this prediction, and is qualitatively consistent with the onset of Pauli blocking.

Comparing the relaxation rate to the gradient in spin density allows us to also measure the spin diffusivity \Ds. At the center of the trap, where the trapping forces vanish, the spin current density $J_s$ is given by the spin diffusion equation~\cite{pari09evap}
$$J_s = -\Ds\frac{\partial (n_{\uparrow} - n_{\downarrow})}{\partial z},$$
where $n_\uparrow(\downarrow)$ is the density of spin up(down) atoms. We calculate the trap-averaged spin current as ${J_s = \frac{1}{2}(n_\uparrow+n_\downarrow) \dot{d}}$, where the densities are evaluated at the trap center.

We find that spin diffusivity is minimum when interactions are resonant (see Supplementary Information). The increase in spin diffusivity for positive scattering length $a$, as well as the  decrease in spin drag, argues against the existence of a ferromagnetic state in repulsive Fermi gases, for which diffusion should stop entirely~\cite{jo09ferro, duin10spin, reca10spin}. Figure 3 reports the measured spin diffusivity as a function of temperature at unitarity. In the high-temperature limit on resonance, one expects $\Ds \sim v/n\sigma \propto T^{3/2}$. At high temperatures, we indeed find this temperature dependence, with a fit giving $\Ds = 5.8(2) \frac{\hbar}{m} (T/T_F)^{3/2}$ for $T/T_F > 2$. Down to our lowest temperatures, the spin diffusivity is seen to attain a limiting value of $6.3(3) \hbar/m$. Fermi liquid theory predicts a rapid increase at low temperatures as $\Ds \propto (T_F/T)^{2}$. From measurements on spin quadrupolar oscillations in highly
spin-imbalanced mixtures it is known~\cite{nasc09imbal} that the resulting crossover from diffusive to ballistic motion occurs below $T/T_F \approx 0.1$.
However, below $T/T_F \approx 0.17$ a balanced gas is superfluid and spin transport possibly inhibited by the superfluid to normal interface~\cite{pari09evap}.

\begin{figure}[ht]
    \begin{center}
    \includegraphics[width=89mm]{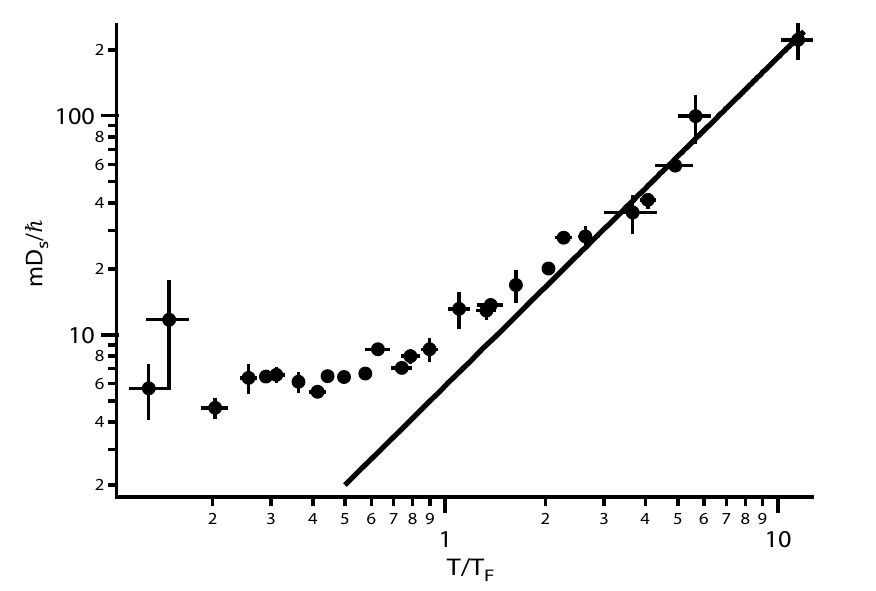}
    \caption[Title]{{\bf Spin diffusivity of a trapped Fermi gas.} Spin diffusivity on resonance (solid circles) as a function of the dimensionless temperature $T/T_F$. At high temperatures, the spin diffusivity obeys the universal $T^{3/2}$ behaviour (solid line).
    At low temperatures the diffusivity approaches a constant value of $6.3(3) \hbar / m$ for temperatures below about $0.5 T_F$, establishing
    the quantum limit of spin diffusion for strongly interacting Fermi gases.}
    \label{fig:spindiffusion}
    \end{center}
\end{figure}

When comparing these results to theoretical calculations, it is important to account for the inhomogeneous density distributions and velocity profiles that result from the trapping potential. For a homogeneous system on resonance, and at high temperatures compared to the Fermi temperature, we predict $\Ds = 1.11 \frac{\hbar}{m} (T/T_F)^{3/2}$ and $\GammaSD=0.90\frac{E_F}{\hbar} (T/T_F)^{-1/2}$ (see Supplementary Information). The measured spin drag coefficient is smaller by a factor of $0.90/0.16(1)=5.6(4)$ while the spin diffusivity is larger by about the same factor, $5.8(2)/1.11 = 5.3(2)$. This factor reflects the inhomogeneity of the system and agrees with an estimate from the Boltzmann transport equation (see Supplementary Information).

Finally, the measured transport coefficients give for the first time access to the temperature-dependence of the spin susceptibility $\chis(T)$ in strongly interacting Fermi gases. Defined as $\chis = \frac{\partial(n_\uparrow-n_\downarrow)}{\partial(\mu_\uparrow -\mu_\downarrow)}$, the spin susceptibility describes the spin response to an infinitesimal effective magnetic field or chemical potential difference $\mu_\uparrow-\mu_\downarrow$ applied to the gas, and is a crucial quantity that can discriminate between different states of matter~\cite{stri09dens}. In a magnetic field gradient, particles with opposite spin are forced apart at a rate determined by the spin conductivity $\sigma_s$, while diffusion acts to recombine them. The balance between the processes of diffusion and conduction therefore determines the resulting magnetization gradient, a connection expressed in the Einstein relation $\chis = \sigma_s/\Ds$~\cite{duin10spin}. In calculating this ratio from observable quantities, the relaxation time $\tau$ cancels, as both $\sigma_s$ and $\Ds$ are proportional to $\frac{1}{\tau}$, yielding
$$\chis = \frac{1}{m\,d\,\omega_z^2}\frac{\partial(n_\uparrow-n_\downarrow)}{\partial z},$$
where $\frac{\partial(n_\uparrow-n_\downarrow)}{\partial z}$ is evaluated near the trap center. The inhomogeneous trapping potential has practically no effect on the measurement of $\chis$, as all quantities involved refer to the vicinity of the center of mass.

Figure 4 reports our findings for the spin susceptibility at unitarity, as a function of the dimensionless temperature $T/T_F$. At high temperatures, we observe the Curie law $\chis = n/T$. In this classical regime of uncorrelated spins, the susceptibility equals the (normalized) compressibility of the gas  $\kappa = \partial n / \partial \mu$ that we also directly obtain from our profiles.
\begin{figure}[ht]
    \begin{center}
    \includegraphics[width=89mm]{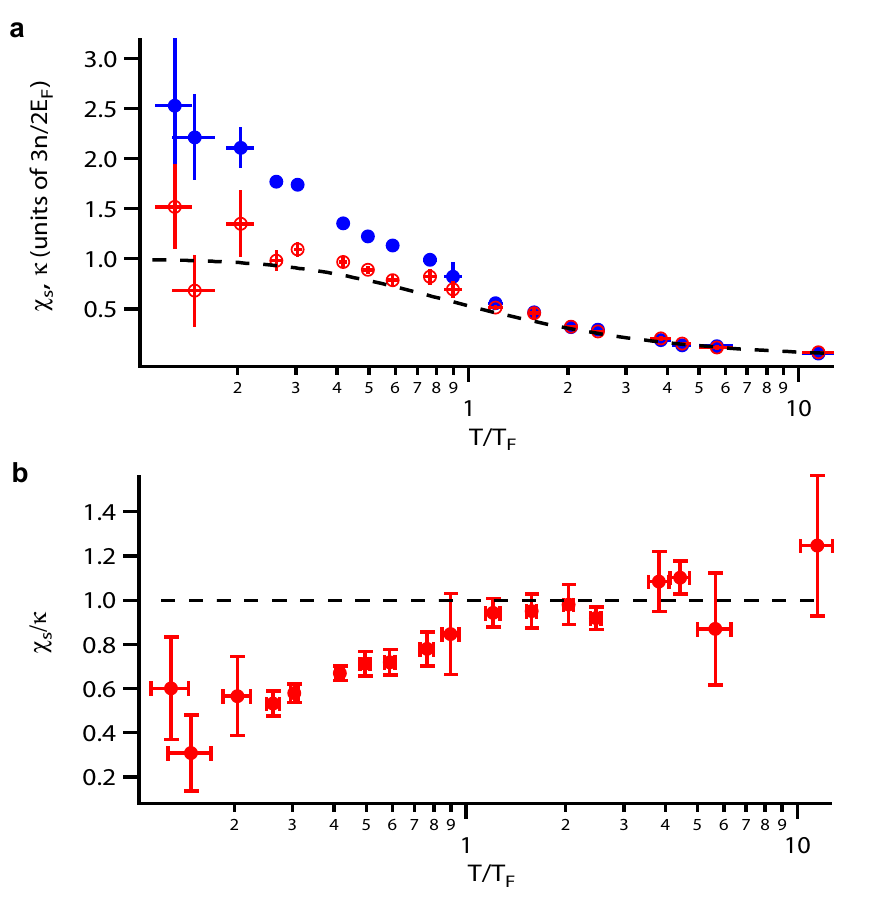}
    \caption[Title]{{\bf Spin susceptibility on resonance determined from the Einstein relation.} {\bf (a)} Compressibility (solid blue circles) and spin susceptibility (open red circles) normalized by the compressibility $\frac{3n}{2E_F}$ of an ideal Fermi gas at zero temperature. For temperatures below the Fermi temperature, the susceptibility becomes suppressed relative to the compressibility, due to interactions between opposite-spin atoms. The spin susceptibility coincidentally matches the compressibility of a non-interacting Fermi gas (dashed line) in the range of temperatures that we could access. {\bf (b)} Red circles: spin susceptibility divided by the compressibility obtained from the same clouds. At temperatures above the Fermi temperature, the ratio of spin susceptibility to compressibility approaches unity (dashed line).}
    \label{fig:spinsusceptibility}
    \end{center}
\end{figure}

At degenerate temperatures, the measured susceptibility becomes smaller than the compressibility. This is expected for a Fermi liquid, where $\chis = \frac{3n}{2 E_F}\frac{1}{1+F_0^a}$ and $\kappa = \frac{3n}{2 E_F}\frac{1}{1+F_0^s}$~\cite{stri09dens} with Landau parameters $F_0^s$ and $F_0^a$ describing the
density (s) and spin (a) response.
The spin susceptibility is expected to strongly decrease at sufficiently low temperatures in the superfluid regime, as pairs will form that will not break in the presence of an infinitesimal magnetic field. For the temperature range of our experiment this behaviour
is not yet observable.
It is currently under debate whether the strongly interacting Fermi gas above the superfluid transition temperature is a Fermi liquid~\cite{nasc10thermo} or a state with an excitation gap (pseudo-gap)~\cite{gaeb10pseudogap,pera10pseudogap}. The opening of a gap in the excitation spectrum would be revealed as a downturn of the spin susceptibility below a certain temperature. Such a downturn is not observed in \chis{} down to $T/T_F \approx 0.2$, and therefore our spin susceptibility data agree down to this point with the expected behavior for a Fermi liquid.

In conclusion, we have studied spin transport in strongly interacting Fermi gases. The spin diffusivity was found to attain a limiting value of about $6.3\hbar/m$, establishing the quantum limit of diffusion for strongly interacting Fermi gases. Away from resonance the diffusivity increases. This casts doubt on the possibility of stabilizing a ferromagnetic gas on the repulsive side of the Feshbach resonance~\cite{jo09ferro}, which would require a vanishing diffusivity~\cite{duin10spin}. The observed slow relaxation of spin excitations is a likely explanation for the surprising -- possibly non-equilibrium~\cite{pari09evap} -- profiles in imbalanced Fermi gases observed at Rice ~\cite{part06phase}, which did not agree with equilibrium measurements at MIT~\cite{zwie05imbalance,shin06phase} and at the ENS ~\cite{nasc09imbal}. Our measurements of the temperature dependence of the spin diffusivity at low temperatures disagree with the expected behavior of a Fermi liquid, while the spin susceptibility that we measure is consistent with a Fermi liquid picture. An interesting subject of further study is whether spins are still able to diffuse through the superfluid, or whether they travel around it, avoiding the superfluid due to the pairing gap.
\\
\\
\textbf{Methods Summary}

The spin mixture is initially prepared at \prepField. To separate the spin components, we ramp the total magnetic field to \kickField, where the magnetic moments of the two spin states are unequal, and apply two magnetic field gradient pulses. We then bring the total magnetic field to the Feshbach resonance in about 2 ms.

To reach low temperatures during the approach to equilibrium, evaporative cooling is applied, at \resField, by gradually lowering the depth of the optical dipole trap. To reach high temperatures, we heat the atoms by switching off the optical dipole trap for up to 3 ms to allow the atoms to expand before recapturing them. We then set the final depth of the dipole trap so that the atom number and the temperature remain nearly constant during the approach to equilibrium.

Spin selective imaging is performed via \textit{in situ} absorption or phase contrast imaging using two 4 $\mu$s imaging pulses separated by 6 $\mu$s. These images give the column densities of each spin state, from which we obtain the three-dimensional density via an inverse Abel transform~\cite{shin06phase}. The gradient in the spin density is obtained from a  linear fit to the polarization versus $z$.

We determine the temperature of the clouds by fitting the density versus potential energy in the vicinity of $z=0$, but for all $r$, to the equation of state of the unitary Fermi gas, measured recently by our group~\cite{ku10equa}. The trapping potential itself is determined by summing the densities of hundreds of clouds, using the known axial, harmonic trapping potential to convert equidensity lines to equipotential lines and fitting the result to an analytic model.
\\
\\
\textbf{Acknowledgements}

We would like to thank Georg Bruun, Chris Pethick, David Huse, and Wilhelm Zwerger for fruitful discussions, and André Schirotzek for help with the early stages of the experiment. This work was supported by the NSF, AFOSR-MURI, ARO-MURI, ONR, DARPA YFA, a grant from the Army Research Office with funding from the DARPA OLE program, the David and Lucille Packard Foundation and the Alfred P. Sloan Foundation.
\\
\\
\textbf{Author Information}

Correspondence and requests for materials should be addressed to A. S.~(email: atsommer@mit.edu).
\bibliography{Spin_Transport_Arxiv2}
\clearpage
\begin{widetext}
\begin{center}
\textbf{Universal Spin Transport in a Strongly Interacting Fermi Gas}\\
\textbf{Supplementary Information}
\end{center}
\section{Dependence on Interaction Strength}
\newcommand{\figbouncevsa}{5}
\newcommand{\figtransportvsa}{6}
\newcommand{\figdensity}{7}
We study the dependence of the spin transport properties of the system on interaction strength by ramping to a variable final field in the vicinity of the Feshbach resonance and measuring the subsequent evolution of the system. Figure \figbouncevsa{} shows the results of colliding two clouds at different fields, revealing the transition from transmission of the clouds through each other to reflection of the clouds as the mean free path becomes smaller than the cloud size. When the scattering length is set to zero (Fig. \figbouncevsa.{\bf a}), the center of mass separation oscillates at the trap frequency $\omega_z$. On resonance (Fig. \figbouncevsa.{\bf g}), the observed oscillation frequency of 1.63(2) $\omega_z$ is intermediate between the frequency 1.55 $\omega_z$ of the axial breathing mode of a unitary Fermi gas in the hydrodynamic limit~\cite{vare06fermi} and the non-interacting value of 2 $\omega_z$, as the system contains a hydrodynamic region at the center, and is non-interacting in the spin-polarized wings. Figure \figtransportvsa{} shows the spin transport coefficients versus interaction strength. The spin drag coefficient exhibits a maximum on resonance (\figtransportvsa.{\bf a}), while the spin diffusion coefficient is minimum on resonance (Fig. \figtransportvsa.{\bf b}).
\begin{figure}[ht]
    \begin{center}
    \includegraphics[width=183mm]{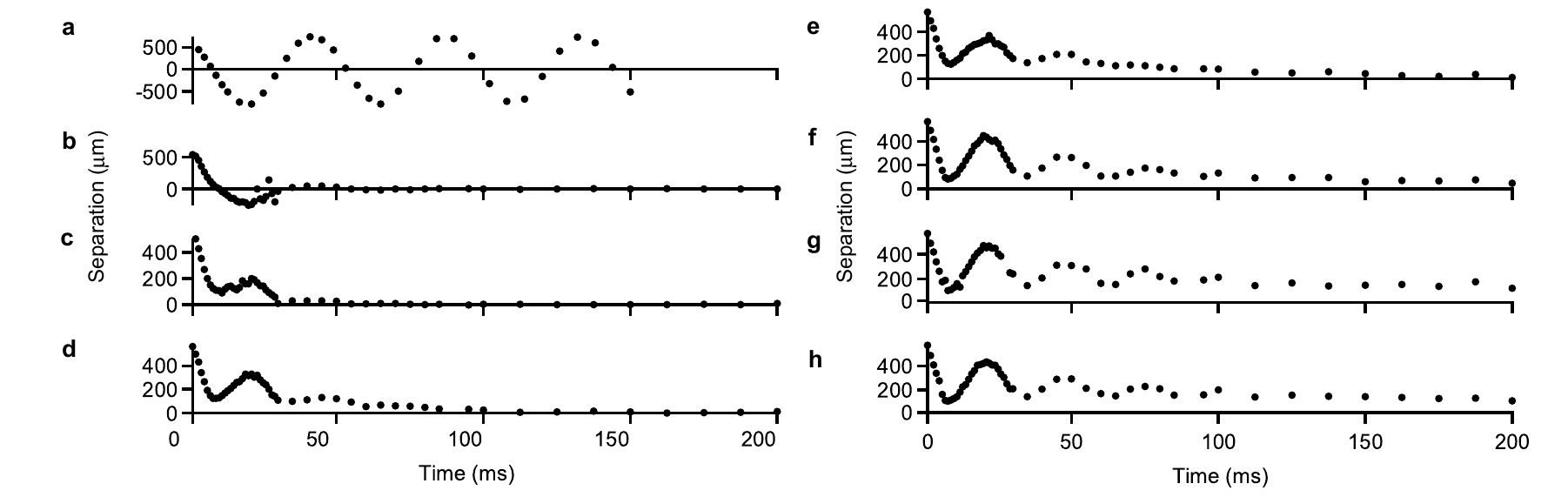}
    \caption[Title]{{\bf Collision between spin up and spin down clouds with varying interaction strength}. After separating the spin components, the magnetic field was ramped to a variable value near the Feshbach resonance to reach different interaction strengths. The interaction parameter $k_F a$, with $k_F = (6\pi^2 n)^{1/3}$ and $n$ the central density per spin component, was determined by averaging the values of $k_F$ obtained from images taken after 200 ms of evolution time (not shown). The values of $k_F a$ were ({\bf a}) 0, ({\bf b}) 0.08, ({\bf c}) 0.13, ({\bf d}) 0.19, ({\bf e}) 0.26, ({\bf f}) 1.2, ({\bf g}) $\infty$, and ({\bf h}) -1.5}
    \label{fig:bouncevsa}
    \end{center}
\end{figure}

\begin{figure}[ht]
    \includegraphics[width=183mm]{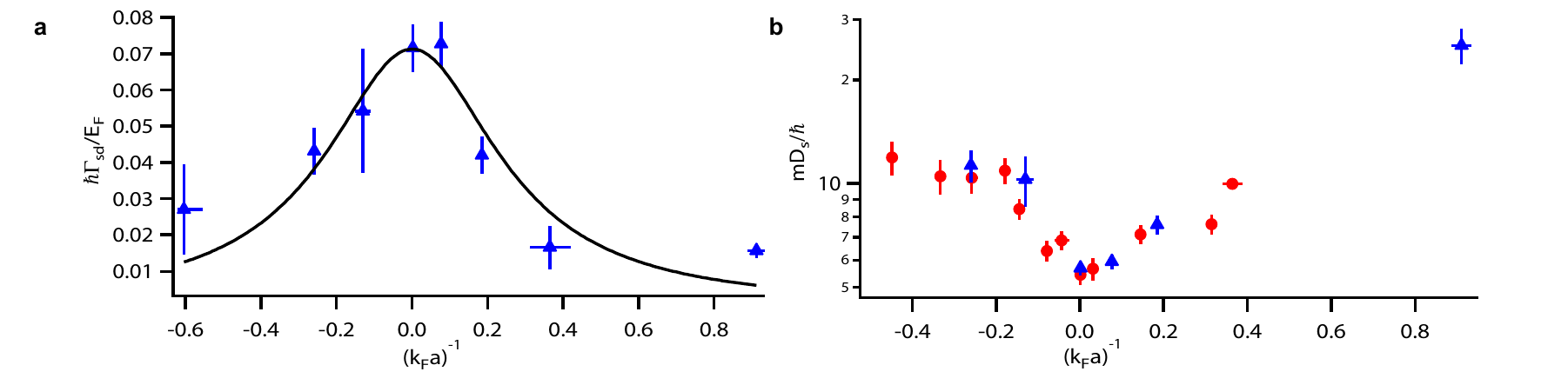}
    \caption[Title]{Spin drag coefficient ({\bf a}) and spin diffusivity ({\bf b}) across the Feshbach resonance. The spin-separated clouds were cooled at \resField{} to $T/T_F\approx0.16$ (blue triangles) and to $T/T_F\approx0.40$ (red cirlces) before ramping to the final field. The largest spin drag and smallest spin diffusivity occur at the Feshbach resonance, where $1/k_F a = 0$. The solid line in ({\bf a}) is a Lorenztian fit. To reduce clutter, the 0.4 $T_F$ data are not shown in ({\bf a}).}
    \label{fig:transportvsa}
\end{figure}

\section{Image Analysis}
The experiment takes place in a trapping potential of the form
\begin{align*}
    V(r,z) = \frac{1}{2}m\omega_z^2 z^2 + V_r(r),
\end{align*}
where $r=\sqrt{x^2 + y^2}$. Here $z$ is the symmetry axis of the trap, and we image along the $y$ axis.

From each run of the experiment we obtain two-dimensional column densities $n^{\rm 2d}_\sigma(x,z)$ of both spin states $\sigma=\uparrow,\downarrow$. Fitting a two-dimensional Gaussian to each column density provides a measurement of the center of mass of each spin state. We subtract the $z$ components of the centers of mass of the two spin states to obtain the separation parameter $d=\left<\zup\right> - \left<\zdown\right>$.
The three-dimensional densities $n_\sigma(r,z)$ are obtained using a numerical implementation of the inverse Abel transformation (similar to \cite{bock61tran}):
\begin{align*}
    n_\sigma(r_i,z) = \sum_{j=i}^{j_{\rm max}} \frac{n^{\rm 2d}_\sigma(x_{j+1},z)-n^{\rm 2d}_\sigma(x_j,z)}{x_{j+1}-x_j} \; \mathrm{ln}\left[\frac{j+1+\sqrt{(j+1)^2-i^2}}{j+\sqrt{j^2-i^2}}\right],
\end{align*}
For each line of constant $z$, the values of $n_\sigma(r_i,z)$ with small $i$ are sensitive to noise in $n^{\rm 2d}_\sigma(x_j,z)$ for small $j$. To reduce noise, we fit a one-dimensional Gaussian to $n^{\rm 2d}_\sigma(x_j,z)$, for each value of $z$ and for $j<4$ (corresponding to the innermost 5 $\mu m$ of the clouds
or less than a tenth of the Fermi-radius), and use this fit (with sub-resolution sampling to reduce discretization error) in the above formula.

The temperature is determined by analyzing the densities $n_\sigma(r,z)$ for $\left|z\right| < 60$ $\mu$m (as the $1/\mathrm{e}$ radii of the clouds range from 100 to 400 $\mu$m, this restriction excludes the more polarized, large $z$, regions of the clouds). We obtain the temperature of each cloud by fitting the density versus potential energy $V$ to
\begin{align*}
    n = \lambda^{-3}f(\beta\mu - \beta V),
\end{align*}
where $\beta = \frac{1}{k_B T}$, $\lambda = \sqrt{\frac{2\pi\hbar^2}{m k_B T}}$, $T$ is the temperature, $\mu$ is the chemical potential, and $f$ is a universal function defining the equation of state of the unitary Fermi gas~\cite{ku10equa}, with $\mu$ and $T$ as the fit parameters.

\section{Measurement of Transport Coefficients}
Near equilibrium, $d$ decays exponentially, and we fit the measured values to
\begin{align*}
    d(t) = d_0 \mathrm{e}^{-t/\tau}.
\end{align*}
The spin drag coefficient is then
\begin{align*}
    \GammaSD = \omega_z^2\tau.
\end{align*}
To non-dimensionalize this quantity we multiply by $\hbar$ and divide by the Fermi energy $E_F = \frac{\hbar^2}{2m}(3\pi^2 n(\vec{0}))^{2/3}$, where $n(\vec{0}) = n_\uparrow(\vec{0})+n_\downarrow(\vec{0})$ is the total three-dimensional density at $(r,z)=(0,0)$.

After fitting to $d(t)$ to obtain $\tau$, we obtain the spin diffusivity from each spin up-spin down image pair as
\begin{align*}
    \Ds = \frac{n(\vec{0}) d}{2 g \tau},
\end{align*}
where $g = \frac{\partial(n_\uparrow-n_\downarrow)}{\partial z}|_{z=0, r=0}$ is the spin density gradient. The spin density gradient is obtained from the slope $b$ of a linear fit to the polarization $p(z) = \frac{n_\uparrow - n_\downarrow}{n_\uparrow + n_\downarrow}$ as a function of $z$ at $r=0$ using the formula
\begin{align*}
    g = b\cdot (n_\uparrow(\vec{0})+n_\downarrow(\vec{0})).
\end{align*}

The Einstein relation provides the spin susceptibility as
\begin{align*}
    \chis &= \sigma_s / \Ds = \frac{n(\vec{0})}{2 m\, \Ds \, \GammaSD}\\
          &= \frac{g}{m\, d\, \omega_z^2}.
\end{align*}
Although derived using transport coefficients, the final expression for $\chis$ does not actually depend on the relaxation time $\tau$, so it gives an independent value of $\chis$ from each run of the experiment.

\section{Theoretical Calculation of Transport Coefficients}
The transport coefficients in the classical limit $T\gg T_F$ may be calculated using the Boltzmann transport equation. The Boltzmann transport equation describes the evolution of the semi-classical distribution functions $f_{\sigma}(\vec{r},\vec{v},t)$, with $\sigma=\uparrow, \downarrow$:
\begin{align*}
    \frac{\partial f_{\sigma}}{\partial t} + \vec{v}\cdot\nabla_{r}f_{\sigma} + \frac{\vec{F}}{m}\cdot\nabla_{v}f_{\sigma} = I_{\rm coll}[f_{\uparrow},_{\downarrow}]
\end{align*}
\\
Define the joint distribution function $f(\vec{r}_\uparrow,\vec{r}_\downarrow,\vec{v}_\uparrow,\vec{v}_\downarrow) = f_{\uparrow}(\vec{r}_\uparrow,\vec{v}_\uparrow) f_\downarrow(\vec{r}_\downarrow,\vec{v}_\downarrow)$. Any quantity of the form $\chi(\vec{r}_\uparrow,\vec{r}_\downarrow,\vec{v}_\uparrow,\vec{v}_\downarrow)$ can be averaged over $f$~\cite{guer99osc}:
\begin{align*}
\left<\chi\right> = \frac{1}{N_\uparrow N_\downarrow} \int {\rm d}^3 r_\uparrow {\rm d}^3 r_\downarrow {\rm d}^3 v_\uparrow {\rm d}^3 v_\downarrow(\chi \, f)\;,
\end{align*}
where $N_\sigma$ is the number of atoms with spin $\sigma$.

We generalize to arbitrary scattering length $a$, and non-uniform drift velocity, the calculation by Vichi and Stringari~\cite{vich99coll} of the equation of motion for the relative coordinates of the two spin components in a harmonic trap of angular frequency $\omega_z$ along the $z$ axis. From the Boltzmann equation,
\begin{align*}
    \partial_t \left<\zup- \zdown\right> &= \left<\vzup-\vzdown\right>\\
    \mathrm{and}\\
   \partial_t \left<\vzup-\vzdown\right> &=  -\left<\vzup-\vzdown\right>\Gamma_{\rm sd} -\omega_z^2 \left<\zup- \zdown\right>,
\end{align*}
where the spin drag coefficient is
\begin{align*}
    \Gamma_{\rm sd} &= \frac{\left<(v_{z\uparrow}-v_{z\downarrow})I_{\rm coll}[f]\right>}{\left<v_{z\uparrow}-v_{z\downarrow}\right>}.
\end{align*}
The full collision integral reads
\begin{align*}
\left<(v_{z\uparrow}-v_{z\downarrow})I_{\rm coll}[f]\right> = - \frac{1}{2}\frac{N_\uparrow + N_\downarrow}{N_\uparrow N_\downarrow}
\int {\rm d}^3r\int \frac{{\rm d}^3 p_\uparrow}{(2\pi\hbar)^3}\int \frac{{\rm d}^3 p_\downarrow}{(2\pi\hbar)^3}\int {\rm d}\Omega \frac{{\rm d}\sigma}{{\rm d}\Omega} \left|\vec{v}_\uparrow - \vec{v}_\downarrow\right| (v_{z\uparrow}-v_{z\downarrow}) \times \\
\left[(1-f_{\uparrow})(1-f_{\downarrow})f_{\uparrow}'f_{\downarrow}' - f_{\uparrow}f_{\downarrow}(1-f_{\uparrow}')(1-f_{\downarrow}')\right].
\end{align*}
The integral describes the scattering of particles $\uparrow$ and $\downarrow$ into new states $\uparrow'$ and $\downarrow'$.
We assume that the distribution functions near equilibrium are $f_\sigma(\vec{r},\vec{v}) = f^0(\vec{r},\vec{v} - \vec{v}_{\sigma,\rm drift})$, where $f^0$ is the equilibrium distribution of both spin states (assumed to be in the classical limit) and $\vec{v}_{\sigma,\rm drift}=\pm v_{\rm d}(\vec{r})\hat{z}$ is the drift velocity of spin $\sigma$. The collision integral can then be calculated (for details, see for example~\cite{smit89transport}) and yields
\begin{align*}
    \Gamma_{\rm sd} &= \frac{16}{3} \sigma(T/T_a) \left(\frac{k_B T}{\pi m}\right)^{1/2}n_0(\vec{0})/\alpha,\\
\end{align*}
with
\begin{align*}
    \sigma(T/T_a) &= 4\pi a^2 \int {\rm d} u \frac{1}{1 + \frac{T}{T_a} u^2} u^5 e^{-u^2},\\
    \alpha &= n_0(\vec{0})\frac{\int  n_0 v_{\rm d} \; {\rm d}^3r}{\int n_0^2 v_{\rm d} \;{\rm d}^3r },
\end{align*}
$n_0=\int f^0\mathrm{d}^3v$ is the equilibrium density of each spin state, and $T_a = \frac{\hbar^2}{k_B m a^2}$ is the temperature scale associated with the scattering length $a$. The limiting values of the temperature-dependent average scattering cross section $\sigma(T/T_a)$ are
\begin{align*}
    \sigma (T \ll T_a) = 4 \pi a^2
\end{align*}
and
\begin{align*}
    \sigma (T \gg T_a) = \frac{2\pi \hbar^2}{m k_B T} = \lambda^2,
\end{align*}
with the thermal de Broglie-wavelength $\lambda$. For unitary interactions, $T_a = 0$, and the latter limit applies. Define the Fermi energy as $E_F = \frac{\hbar^2}{2m}(6\pi^2 n_0(\vec{0}))^{2/3}$. In the unitary limit where $\sigma=\lambda^2$ we find
\begin{align*}
    \frac{\hbar\GammaSD}{E_F} = \frac{32\sqrt{2}}{9\pi^{3/2}\alpha}\sqrt{\frac{E_F}{k_B T}}
                                =\frac{0.90}{\alpha}\sqrt{\frac{E_F}{k_B T}}.
\end{align*}
For a uniform system $\alpha = 1$, while for a harmonically trapped system with a uniform drift velocity we find $\alpha=2^{3/2}$. However, the drift velocity profile cannot be uniform: Even if it started out uniform, spin currents would get damped faster in the center of the overlap region of the two clouds, where the collision rate is high, than in the wings, where it is low. A non-uniform drift velocity profile will develop. The nature of the linearized Boltzmann equation allows for a variational principle where trial functions replacing the true distribution function $f_\sigma(\vec{r},\vec{v})$ yield upper bounds on the actual spin drag coefficient~\cite{smit89transport}. Minimizing $\GammaSD$ for a trial class of non-uniform drift velocities $v_{\rm d}(\vect{r}) = v_{d0}(1-n_0(\vect{r})^\gamma/n_0(\vect{0})^\gamma)$ yields $\gamma \rightarrow 0$. In this case $\alpha=2^{5/2}\approx5.7$.

The spin diffusivity measured in our experiment for $T\gg T_F$ is then
\begin{align*}
    \Ds &=\frac{1}{\chis}\frac{n_0(\vec{0})}{m\GammaSD}\\
                &= \frac{9\pi^{3/2}\alpha}{32\sqrt{2}} \frac{\hbar}{m} \left(\frac{T}{T_F}\right)^{3/2} \\
                &=1.1\alpha \frac{\hbar}{m} \left(\frac{T}{T_F}\right)^{3/2}.
\end{align*}
The bulk value for $\alpha = 1$ represents a lower bound on the diffusivity. It is known from similar calculations that in the high-temperature limit, this lower bound should be within a few percent of the actual bulk value~\cite{smit89transport}. The effect of inhomogeneities in a harmonic, cylindrically symmetric trap increases the diffusivity by $\alpha \approx 5.7$.
The high-temperature result for the diffusivity has been obtained independently by Georg Bruun~\cite{bruu10spin}.

\end{widetext}
\end{document}